\documentclass[conference]{IEEEtran}
\IEEEoverridecommandlockouts
\usepackage{cite}
\usepackage{amsmath,amssymb,amsfonts}
\usepackage{balance}
\usepackage{algorithm}

\usepackage{graphicx}
\usepackage[inline]{enumitem} 

\usepackage{textcomp}
\usepackage{verbatim}
\usepackage{algorithm,algpseudocode}
\usepackage{mathtools}
\usepackage{graphicx}
\usepackage{mathrsfs}
\usepackage{amsmath}
\usepackage{amsfonts}
\usepackage{amssymb}
\newcommand{\CLASSINPUTbottomtextmargin}{0.95in}
\usepackage{caption}
\usepackage{xcolor}
\def\BibTeX{{\rm B\kern-.05em{\sc i\kern-.025em b}\kern-.08em
    T\kern-.1667em\lower.7ex\hbox{E}\kern-.125emX}}

\makeatletter
\newsavebox{\@brx}
\newcommand{\llangle}[1][]{\savebox{\@brx}{\(\m@th{#1\langle}\)}%
  \mathopen{\copy\@brx\kern-0.5\wd\@brx\usebox{\@brx}}}
\newcommand{\rrangle}[1][]{\savebox{\@brx}{\(\m@th{#1\rangle}\)}%
  \mathclose{\copy\@brx\kern-0.5\wd\@brx\usebox{\@brx}}}
\makeatother

\begin{document}

\title{Fast Geometric Learning of MIMO Signal Detection over Grassmannian Manifolds\vspace{-0.1in}\\
\thanks{This research was supported by the Center
for Assured and Resilient Navigation in Advanced TransportatION Systems
(CARNATIONS) under the US Department of Transportation (USDOT)’s University
Transportation Center (UTC) program (Grant No. 69A3552348324).}
}

\author{\IEEEauthorblockN{Rashed Shelim\IEEEauthorrefmark{1}\IEEEauthorrefmark{2}, Walid Saad\IEEEauthorrefmark{1}, and Naren Ramakrishnan\IEEEauthorrefmark{2}}
\IEEEauthorblockA{\IEEEauthorrefmark{1}Department of Electrical and Computer Engineering, Virginia Tech, USA 
 \\
 \IEEEauthorrefmark{2}Department of Computer Science, Virginia Tech, Arlington, VA 22203\\
Emails: \{rasheds@vt.edu, walids@vt.edu, naren@vt.edu\}\vspace{-0.06in}}
}

\maketitle
\begin{abstract} Domain or statistical distribution shifts are a key staple of the wireless communication channel, because of the dynamics of the environment. Deep learning (DL) models for detecting multiple-input multiple-output (MIMO) signals in dynamic communication require large training samples (in the order of hundreds of thousands to millions) and online retraining to adapt to domain shift. Some dynamic networks, such as vehicular networks, cannot tolerate the waiting time associated with gathering a large number of training samples or online fine-tuning which incurs significant end-to-end delay. In this paper, a novel classification technique based on the concept of geodesic flow kernel (GFK) is proposed for MIMO signal detection. In particular, received MIMO signals are first represented as points on Grassmannian manifolds by formulating basis of subspaces spanned by the rows vectors of the received signal. Then, the domain shift is modeled using a geodesic flow kernel integrating the subspaces that lie on the geodesic to characterize changes in geometric and statistical properties of the received signals. The kernel derives low-dimensional representations of the received signals over the Grassman manifolds that are invariant to domain shift and is used in a geometric support vector machine (G-SVM) algorithm for MIMO signal detection in an unsupervised manner. Simulation results reveal that the proposed method achieves promising performance against the existing baselines like OAMPnet and MMNet with
only $1,\!200$ training samples and without online retraining.
\end{abstract}


\section{Introduction}
Signal detection under rapid domain shift (e.g., distribution shift) is one of the fundamental problems in dynamic wireless networks, such as vehicular networks, in which a rapid domain shift occurs frequently due to different channel impairments such as Doppler shift and multi-path fading.  Mobility is a main culprit behind much of the dynamics in such networks. The presence of such mobility and dynamics creates challenges for multiple-input multiple-output (MIMO) signal detection. Conventional MIMO signal detection algorithms such as maximum likelihood detector (MLD), minimum mean square
error (MMSE), 
zero-forcing (ZF) \cite{ZF}, and approximate message passing (AMP) \cite {AMP} are model-driven, and require the knowledge of perfect channel state information (CSI). However, it is difficult to obtain accurate instantaneous CSI in a highly dynamic environment due to high mobility. Consequently, the conventional algorithms designed in \cite{ZF, AMP} perform poorly under mobility, or they are too complex to be practical, as they treat the acquired imperfect CSI as if it were perfect. 
 
Machine learning techniques~\cite{x_1,x_2,AGI} can be a promising approach to address these challenges, as shown in \cite{DetNet,OAMPnet, HyperMIMO, MMNet}. Indeed, several deep learning (DL) based approaches were proposed in \cite{DetNet,OAMPnet, HyperMIMO, MMNet} for MIMO signal detection, and they are shown to outperform conventional MIMO signal detection algorithms proposed in \cite{ZF, AMP}. However, the prior art in \cite{DetNet, OAMPnet, HyperMIMO, MMNet} relies on highly parameterized deep neural networks (DNN) models so as to represent a broad range of mappings for various channel realizations. As such, massive training samples in the order of hundreds of thousands to millions are typically required by the works in \cite{DetNet, OAMPnet, HyperMIMO, MMNet} in order to obtain a desirable mapping. Meanwhile, the dynamic nature of wireless channels, particularly in presence of mobility,  causes a rapid domain shift for the communication signals.  In such dynamic and mobile environment, DNNs trained for a given channel may no longer perform well on future unseen channel realizations. Thus, online training or fine-tuning is required by techniques like those in \cite{DetNet, OAMPnet, HyperMIMO, MMNet} in order to adapt to domain shifts. Dynamic settings, such as vehicular networks, cannot tolerate the waiting time associated with gathering a large number of training samples and online retraining, which incur significant end-to-end delay. In such settings if the information is time-critical (e.g., navigation in autonomous vehicles), the delay experienced by the system in order to adapt to new domains could lead to dire consequences. Clearly, there is a lack of existing approaches for MIMO signal detection that can overcome the grand challenge of enabling a wireless system to effectively adapt to dynamic environments, without the need for massive volumes of training samples and frequent online retraining.

The main contribution of this paper is a novel MIMO signal detection method that exploits a geodesic flow kernel (GFK)~\cite{GFK_new} for the purpose of learning domain shifts over Grassmannian manifolds. The GFK is particularly used in geometric support vector machine (G-SVM) for detecting MIMO signals in an unsupervised manner.  Geometric approaches, including Grassmannian and Riemannian geometry, have been used to address challenges in wireless communication systems, such as the design of beamforming codebook in \cite {beamforming}, wireless resource allocations \cite{paper1,paper2,paper3} and deriving the capacity of the noncoherent multiple-antenna channel \cite{tse}. However, the works in \cite {beamforming, paper1,paper2,paper3,tse} do not consider the Grassmannian geometry to study the MIMO signal detection under domain shift.

In our approach, we first represent the received MIMO signals as points on Grassmanninan manifold by performing principal component anaylsis (PCA) \cite{PCA} which identifies the subspaces where the variance of the received signals are maximized. Then, we show that the channel variations due to mobility flows along a geodesic between training and test data point which belong to different domains. The flow along the geodesic represents incremental changes in geometric and statistical properties of the received signals due to mobility. The geodesic flow kernel derives low-dimensional representations of the received MIMO signals over the Grassman manifold that are invariant to domain shift and used in G-SVM for MIMO signal detection. 
Simulation results shows that the proposed method achieves promising performance compared to the state-of-the-art baselines like OAMPNet \cite{OAMPnet}, HyperMIMO \cite{HyperMIMO} and MMNet\cite{MMNet} with
only $1200$ training samples and without online training.

The rest of this paper is organized as follows. The system model and problem formulation are presented in Section~\ref{system_model1}.  Section~\ref{solution} presents the proposed learning-based MIMO signal  detection method. Simulation results are presented in Section~\ref{simulation}. Finally, conclusions are drawn in Section~\ref{conclusion}.

\section{System Model and Problem Formulation}\label{system_model1}
\subsection{Preliminaries} 
A Grassmannian manifold $\mathsf{G}_{T,N}$ is a set of $N$-dimensional subspaces in $\mathbb{C}^T$ that can be used to define a smooth Riemannian manifold with geometric, differential and probabilistic structure, where $T$ is the dimensionality of the original data \cite{Grassman_paper}. A point $\mathsf{P}$ on $\mathsf{G}_{T,N}$ is a subspace and is typically represented by a basis matrix spanning the subspace. 
The subspace can be computed by utilizing PCA \cite{PCA}. PCA computes a subspace on the Grassmannian manifold by extracting the principal components (i.e., eigenvectors) from the covariance matrix of the input data. These principal components represent the most significant directions of variation in the data and form the basis for the subspace within the Grassmannian manifold.

\subsection{System Model}
We consider a high mobility uplink MIMO scenario consisting of $M$ transmit antennas at mobile node (e.g., ground or aerial vehicle) and $N$ receive antennas at the receiver (e.g., edge server). The mobile node transmits one out of $Z$ possible encoded symbols $\boldsymbol{s}^{(i)} \in \mathbb{C}^{1\times L}$ with $i \in \mathcal{Z}= \{1,2,\ldots,Z\}$. 
Each symbol $\boldsymbol{s}^{(i)}$ is a complex random variable with $L$ being an integer multiple of $M$. 
Then, $\boldsymbol{s}^{(i)}$ can be divided into $1 \times T$ sub-vectors with $T=L/M$, i.e., $\boldsymbol{s}^{(i)}=[\boldsymbol{s}^{(i)}_1,\boldsymbol{s}^{(i)}_2,\ldots,\boldsymbol{s}^{(i)}_{M}]$. Thus, the transmitted symbol $\boldsymbol{s}^{(i)}$ can be converted into a matrix $\boldsymbol{X}^{(i)}\in \mathbb{C}^{M \times T}$. In situations where  $L$ is not an integer multiple of $M$, zero-padding can be used to extend $\boldsymbol{s}^{(i)}$ to meet the integer-multiple constraint. Matrix $\boldsymbol{X}^{(i)}\in \mathbb{C}^{M \times T}$ is transmitted by the mobile node over $T$ slots through $M$ antennas.

\begin{figure}[htbp]
\centerline{
\includegraphics[width=7.9cm, height=7.25cm]{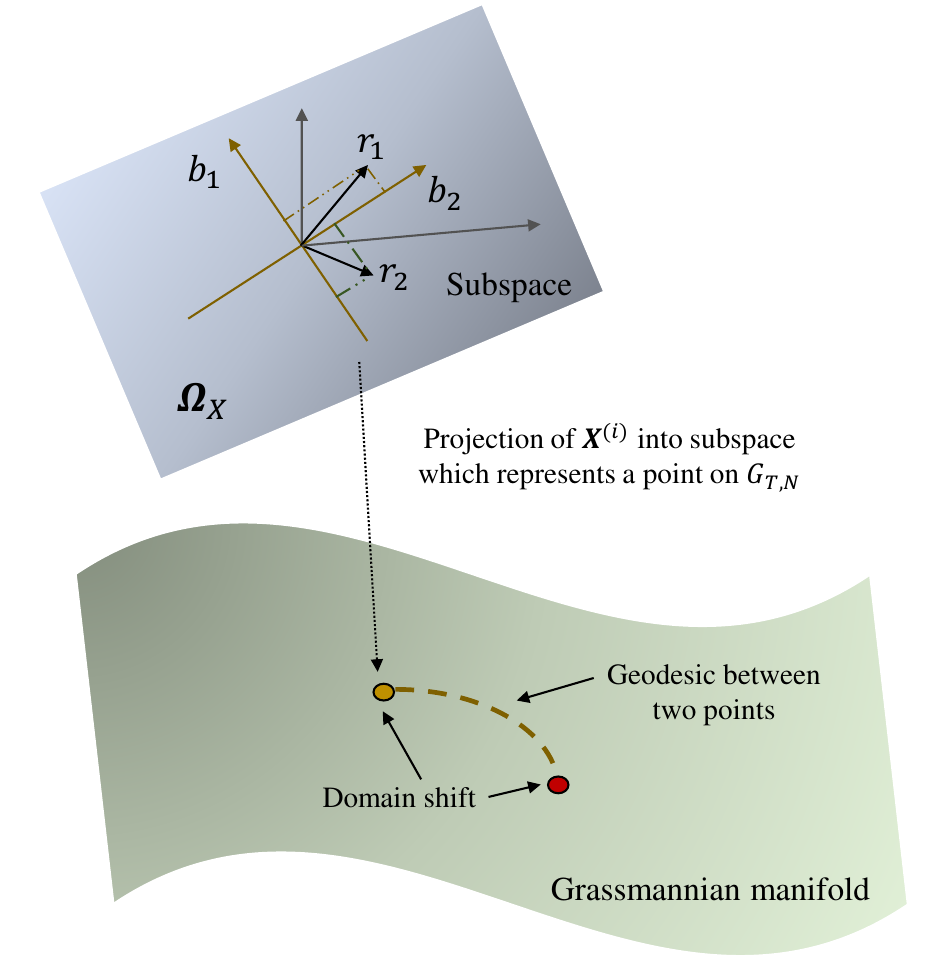}}
\caption{\small{An example of the transformation of coordinate system through the representation of data on Grassmannian manifolds. Here, $[b_1,b_2]$ is the basis of $\Omega_{\boldsymbol{X}}$, $r_1, r_2$ are the row vector of $\boldsymbol{X}$}. $\boldsymbol{U}_{\boldsymbol{X}}=[u_{i,j}]$ where $u_{i,j}$ is the length of the component of $r_i$ in the direction of $b_j$.}
\label{coordinate}
\end{figure}
High mobility leads to temporally-correlated MIMO channels. Under the assumption of rich scattering, classical Clark's model \cite{clark} is used to quantify the level of channel temporal correlation associated with the node speed. We divide time into baseband sampling
intervals, called time slots. Then, the realization of the slot-$t$ MIMO channel from the mobile node to the receiver can be represented by $\boldsymbol{H}_t\!\in\!\mathbb{C}^{N\times M}$.
To reflect that channel's temporal
variation, the transmitted symbol $\boldsymbol{X}^{(i)}$ can be written in terms of its columns, i.e., $\boldsymbol{X}^{(i)}=[\boldsymbol{x}^{(i)}_1,\boldsymbol{x}^{(i)}_2,\ldots,\boldsymbol{x}^{(i)}_T]$. Then, the received signal at the receiver due to the transmission of $\boldsymbol{X}^{(i)}$ can be represented by matrix $\boldsymbol{Y}^{(i)}=[\boldsymbol{y}^{(i)}_1,\boldsymbol{y}^{(i)}_2,\ldots,\boldsymbol{y}^{(i)}_T]\in \mathbb{C}^{N\times T}$ with 
\begin{align}
    \boldsymbol{y}^{(i)}_t = \sqrt{P} \boldsymbol{H}^{(i)}_t\boldsymbol{x}^{(i)}_t + \boldsymbol{w}^{(i)}, \label{received_symbol}
\end{align}
where $P$ is the transmit power and $\boldsymbol{w}^{(i)}$ is the additive white Gaussian noise (AWGN) vector whose entries are assumed to be mutually independent with zero mean and variance $\gamma_w$, i.e., $w_i\sim \mathcal{N}(w_i,0,\gamma_w^{-1})$. From \eqref{received_symbol}, we observe that due to high mobility, the channel $\boldsymbol{\mathcal{H}^{(i)}}=\{\boldsymbol{H}^{(i)}_t\}_{t=1}^T$ varies throughout the transmission duration of a single symbol across $T$ time slots. 

The transmitted signal $\boldsymbol{X}^{(i)}\!\!\in\!\!\mathbb{C}^{M \times T}$ can be represented as a point $\Omega_{\boldsymbol{X}}^{(i)}$ on  $\mathsf{G}_{T,M}$  spanned by its row vector using PCA, alongh with a matrix $\boldsymbol{U}_{\boldsymbol{X}}^{(i)}\in \mathbb{C}^{M \times M}$ which specifies the $M$ row vectors of $\boldsymbol{X}^{(i)}$ with respect to a canonical basis in $\Omega_{\boldsymbol{X}}^{(i)}$ \cite{tse}. 
The representation of the transmitted signal over Grassmannian manifolds  $\boldsymbol{X}^{(i)} \rightarrow (\boldsymbol{U}_{\boldsymbol{X}}^{(i)},\Omega_{\boldsymbol{X}}^{(i)})$
is the transformation of coordinate system $\mathbb{C}^{M\times T} \rightarrow \mathbb{C}^{M\times M} \times \mathsf{G}_{T,M}$, as shown in Fig.~\ref{coordinate}. To understand the motivation of such representation, we first consider an extreme case without additive noise $\boldsymbol{W}^{(i)}$ such that $\boldsymbol{Y}^{(i)}\!=\!\boldsymbol{\mathcal{H}}^{(i)}\boldsymbol{X}^{(i)}$. In this case, the row vectors of the received signals span the same subspace as $\boldsymbol{X}^{(i)}$, i.e., $\Omega_{\boldsymbol{\mathcal{H}}\boldsymbol{X}}^{(i)}=\Omega_{\boldsymbol{X}}^{(i)}$. This shows that the channel $\boldsymbol{\mathcal{H}}^{(i)}$ affects the transmitted signal $\boldsymbol{X}^{(i)}$ by changing $\boldsymbol{U}_{\boldsymbol{X}}^{(i)}$. 
In consequence, the position of the subspace of the transmitted signal shifts towards the direction of the change in $\boldsymbol{U}_{\boldsymbol{X}}^{(i)}$ over Grassmannian manifolds. For a channel with additive noise, $\boldsymbol{U}_{\boldsymbol{X}}^{(i)}$ is affected by both noise and channel and the $\Omega_{\boldsymbol{X}}^{(i)}$ is affected by noise only. In the high signal-to-noise-ratio (SNR) regime, the impact of noise is negligible compared to the variation in the channel. In this case, it can be intuitively assumed that $\boldsymbol{U}_{\boldsymbol{X}}^{(i)}$ is affected only by the channel~\cite{tse}. 

\subsection{Problem Formulation} \label{problem_formulation}
Let $\mathcal{Y}$ be input space of a dataset $\mathcal{D} = \{(Y_j|P_{\mathsf{Y}}(Y_j)\}_{j=1}^{K}$, where $Y_j\in \mathcal{Y}$ is the input $j$ and $K$ is the total size of the dataset. When $Y_j$ is seen as realization of random variables $\mathsf{Y}$, then, it is possible to define a domain as its marginal distribution $P_{\mathsf{Y}}(Y)$. If the domains of the training $\mathcal{D}_{\textrm{r}}$ and test $\mathcal{D}_{\textrm{s}}$ datasets are different, then $P_{\mathsf{Y}}^s(Y)\neq P_{\mathsf{Y}}^t(Y)$. In other words, the marginal distributions are different. From \eqref{received_symbol}, this domain shift is associated with the distribution of the received signal which is caused by channel variation due to mobility. 

To illustrate how representation over Grassmannian manifolds can assist in addressing the challenge of rapid domain shift under mobility, 
consider two subspaces are computed from the training and test dataset corresponding to its domain $\mathcal{D}_{\textrm{r}}$ and $\mathcal{D}_{\textrm{s}}$, respectively. Those two subspaces are represented as two points on Grassmannian manifolds. For a static channel and zero noise, the two points are identical on $\mathsf{G}_{T,M}$. In this case, the two domains are similar to each other, i.e., the training and test dataset are similarly distributed. In the presence of channel variation and noise, they are two different points on $\mathsf{G}_{T,M}$ and hence, the two domains are far apart on the manifolds. 

In real-world applications, it is very difficult to ensure that the training and test data are drawn from a similar domain. This problem is particularly challenging in dynamic wireless communication settings in which a rapid domain shift occurs frequently due to mobility. 
Hence, our overarching goal is to overcome the challenge of the domain shift problem by identifying meaningful intermediate subspaces computed from the training and the test dataset that are robust to channel variation due to mobility. Specifically, we intend to use these meaningful intermediate subspaces to predict the transmitted MIMO symbol that belongs to domain $\mathcal{D}_{\textrm{s}}$ (i.e., under unseen channel realizations due to mobility) by only using the labeled training dataset that belongs to domain $\mathcal{D}_{\textrm{r}}$.  
\section{GFK based Model for MIMO Signal Detection}\label{solution}
As explained next, the proposed framework consists of the following steps: i) formulation of a geodesic flow curve connecting the training and test domains on the Grassmannian manifolds, ii) computing the GFK, and iii) use of the GFK in G-SVM classifier for MIMO signal detection. 

\subsection{Geodesic Flow for Domain Shift Adaptation}
Since the symbol $\boldsymbol{s}^{(i)}\!\in\!\mathbb{C}^{1\times L}$ is converted into matrix $\boldsymbol{X}^{(i)}\in \mathbb{C}^{M \times T}$ when transmitted over $M$ antennas, it is first converted back to its vector form, i.e.,  $\boldsymbol{\hat{s}}^{(i)}\!\in\!\mathbb{C}^{1\times L'}$, where $L'\!\!=\!\!NT$ and $L'\!\!=\!\!L$ if $M\!=\!N$. This conversion is done by the concatenation of the row vectors of the received signal matrix $\boldsymbol{Y}^{(i)}\in \mathbb{C}^{N \times T}$. Then, PCA is applied to identify the subspace of the received symbol. Since the number of eigenvectors of the received signal is $N$, it is sufficient
to represent a subspace through its orthonormal basis $\boldsymbol{S} \in \mathbb{C}^{L'\times N}$  by applying PCA, where $L'$ is the dimension of the data (i.e., symbol), and $N$ is the dimension of the subspace. To obtain meaningful intermediate subspaces between the training and test domains for handling a domain shift, a set of tools is required that is consistent with the geometry of the space spanned by these $L' \times N$ subspaces.

Let $\boldsymbol{S}_{\textrm{r}}, \boldsymbol{S}_{\textrm{s}}\in \mathbb{C}^{L'\times N}$ be two sets of orthonormal basis of the subspaces obtained by PCA from the received signals in the training and test domains. We define 
$\boldsymbol{R}_{\textrm{r}}\in \mathbb{C}^{L'\times (L'-N)}$ as the orthonormal complement to $\boldsymbol{S}_{\textrm{r}}$, namely
$\boldsymbol{R}_{\textrm{r}}^{\dagger}\boldsymbol{S}_{\textrm{r}}=0$, where $\dagger$ is the Hermitian or matrix conjugate transpose. Using a canonical
Euclidean metric for the Riemannian manifold in a Grassmannian manifold, the geodesic flow can be parametrized as $\mathbf{\Phi}(q):\!\!q\!\in\![0,1]\!\rightarrow\!\mathbf{\Phi}(q)\in \mathsf{G}_{L',N}$ with the constraint $\mathbf{\Phi}(0)=\boldsymbol{S}_{\textrm{r}}$ and $\mathbf{\Phi}(1)=\boldsymbol{S}_{\textrm{s}}$. Then, the geodesic flow that emanates from $\boldsymbol{S}_{\textrm{r}}$ to $\boldsymbol{S}_{s}$ will be given by \cite{GFK_new} 
\begin{align}
    \mathbf{\Phi}(q) &= \boldsymbol{S}_{\textrm{r}}\boldsymbol{U}_1\bold{\Gamma}(q) - \boldsymbol{R}_{\textrm{r}}\boldsymbol{U}_2\bold{\Sigma}(q), \nonumber\\
    &= \begin{bmatrix}\boldsymbol{S}_{\textrm{r}} & \boldsymbol{R}_{\textrm{r}} \end{bmatrix} \begin{bmatrix}
    \boldsymbol{U}_1 & 0 \\
     0 & \boldsymbol{U}_2    
    \end{bmatrix}
    \begin{bmatrix}
        \bold{\Gamma}(q)\\\bold{\Sigma}(q) \label{gflow}
    \end{bmatrix},
\end{align}
where $\boldsymbol{U}_1\in \mathbb{C}^{N\times N}$ and $\boldsymbol{U}_2\in \mathbb{C}^{(L'-N)\times N}$ are orthonormal matrices
given by singular value decomposition (SVD) as follows:
\begin{align}
    \boldsymbol{S}_{\textrm{r}}^{\dagger}\boldsymbol{S}_{\textrm{s}}= \boldsymbol{U}_1\bold{\Gamma}\boldsymbol{V}^{\dagger}, \;\;\;
    \boldsymbol{R}_{\textrm{r}}^{\dagger}\boldsymbol{S}_{\textrm{s}}= -\boldsymbol{U}_2\bold{\Sigma}\boldsymbol{V}^{\dagger},
\end{align}
where $\bold{\Gamma}\in \mathbb{C}^{N\times N}$ and $\bold{\Sigma}\in \mathbb{C}^{N\times N}$ are diagonal matrices. The diagonal elements $\cos{\phi}_l$ and $\sin{\phi}_l$ for $i=1,2,\ldots,N$ are called principle angles i.e., $0\leq\phi_1\ldots\leq\phi_{N}\leq\pi/2$, and are the measure of degree of overlap between $\boldsymbol{S}_{\textrm{r}}$ and $\boldsymbol{S}_{\textrm{s}}$. Based on \eqref{gflow}, the geodesic flow can be viewed as a collection of infinite meaningful subspaces gradually varying from the training domain $\mathcal{D}_{\textrm{r}}$ to the test domain $\mathcal{D}_{\textrm{s}}$ with the variations being caused by the channel dynamics due to mobility. In other words, the channel variations flow along a geodesic between training and test data
points which belong to different domains.
\subsection{Computation of the Geodesic Flow Kernel}
Let $\boldsymbol{\hat{s}}^{(i)}_{\textrm{r}}$ and $\boldsymbol{\hat{s}}^{(i)}_{\textrm{s}}$ be the two versions of the originally transmitted symbol $\boldsymbol{s}^{(i)}$ that are received, respectively, during the training and test phases. Those received symbols belong to different domains. We compute their projections on the geodesic $\mathbf{\Phi}(q)$, i.e., $\mathbf{\Phi}(q)^{\dagger}\boldsymbol{\hat{s}}^{\dagger}$, for a continuous $q$ starting from $0$ to $1$, where $0$ and $1$ indicate the training domain and test domain, respectively. By using projection into all subspaces along the geodesic, we utilize a similarity measure that is robust to channel variations. 
In other words, the net effect of projections is a representation over Grassmannian manifolds that is invariant to the domain shift. All the projections are concatenated into infinite-dimensional feature vectors $\boldsymbol{z}_{\textrm{r}}^{\infty}$ and $\boldsymbol{z}_{\textrm{s}}^{\infty}$. Computationally, we do not need to compute infinitely many projections to define GFK. Instead, the GFK can be defined by inner products between $\boldsymbol{z}_{\textrm{r}}^{\infty}$ and $\boldsymbol{z}_{\textrm{s}}^{\infty}$ as:
\begin{align}
    \langle \boldsymbol{z}_{\textrm{r}}^{\infty}, \boldsymbol{z}_{\textrm{s}}^{\infty} \rangle\!&=\!\int_{0}^{1}\!\!{\Big(\! \mathbf{\Phi}(q)^{\dagger}\big(\boldsymbol{\hat{s}}^{(i)}_{\textrm{r}}\big)^{\dagger}  \Big)}^{\dagger}\!\big( \mathbf{\Phi}(q)^{\dagger}{\big(\boldsymbol{\hat{s}}^{(i)}_{\textrm{s}}\big)}^{\dagger}  \big) dq \nonumber\\&= \boldsymbol{\hat{s}}^{(i)}_{\textrm{r}}\boldsymbol{F}{\big(\boldsymbol{\hat{s}}^{(i)}_{\textrm{s}}\big)}^{\dagger},\label{inner_product} 
\end{align}
where $\boldsymbol{F}\in \mathbb{C}^{L'\times L'}$ is a positive definite matrix 
\begin{align}
    \boldsymbol{F} = \int_0^1 \mathbf{\Phi}(q)\mathbf{\Phi}(q)^{\dagger} dq.
\end{align}
The operation performed in \eqref{inner_product} is the  \textit{kernel trick} \cite{GFK_new}, where a kernel function induces inner products between infinite-dimensional features without explicitly computing an infinite number of projections. Then, the geodesic flow kernel $\boldsymbol{F}$ can be formulated in closed form as follows \cite{GFK_new}:
\begin{align}
      \boldsymbol{F} = \begin{bmatrix}
          \boldsymbol{S}_{\textrm{r}}\boldsymbol{U}_1 \!&\! \boldsymbol{R}_{\textrm{r}}\boldsymbol{U}_2
      \end{bmatrix}
       \begin{bmatrix}
    \bold{\Upsilon}_1 & \bold{\Upsilon}_2 \\
     \bold{\Upsilon}_2 & \bold{\Upsilon}_3
    \end{bmatrix}
    \begin{bmatrix}
    \boldsymbol{U}_1^{\dagger}\!\boldsymbol{S}_{\textrm{r}}^{\dagger}\\
     \boldsymbol{U}_2^{\dagger}\!\boldsymbol{R}_{\textrm{r}}^{\dagger}
    \end{bmatrix},\label{gfk}
\end{align}
where $\bold{\Upsilon}_1$,$\bold{\Upsilon}_2$, and $\bold{\Upsilon}_3$ are diagonal matrices, and the diagonal entries of these matrices are given by \cite{GFK_new}
\begin{align}
    \sigma_{1j}&= \int_0^1 \cos^2(q\phi_j)\;dq= 1\!+\!\frac{\sin{(2\phi_j)}}{2\phi_j},\\ \sigma_{2j}&= \int_0^1 \cos(q\phi_j)\sin(q\phi_j)\;dq =\frac{\cos{(2\phi_j)}-1}{2\phi_j}, \\
    \sigma_{3j}&= \int_0^1 \sin^2(q\phi_j)\;dq =1\!-\!\frac{\sin{(2\phi_j)}}{2\phi_j},
\end{align}
respectively, and $j=1,2,\ldots,N$. 
\vspace{-0.05in}
\subsection{GFK G-SVM for Signal Detection}
Now, let $\{\mathcal{S}_{\textrm{r}},\mathcal{T}_{\textrm{r}}\}\!=\!{\{(\boldsymbol{\hat{s}}_{\textrm{r}}^l,\tau_{\textrm{r}}^l)\}}_{l=1}^{n_{\textrm{r}}}$, with $\boldsymbol{\hat{s}}_{\textrm{r}}^l\in \mathbb{C}^{1\times L'}$, and $\tau_{\textrm{r}}^l \in \mathcal{Z}$ be the set of $n_{\textrm{r}}$ labeled training dataset corresponding to $Z$ classes (i.e., $Z$ possible symbols in alphabet $\mathcal{Z}$) from domain $\mathcal{D}_{\textrm{r}}$. We also define $\{\mathcal{S}_{\textrm{s}}\}={\{\boldsymbol{\hat{s}}_{\textrm{s}}^k\}}_{k=1}^{n_{\textrm{s}}}$, $\boldsymbol{\hat{s}}_{\textrm{s}}^k\in \mathbb{C}^{1\times L'}$ as the set of $n_{\textrm{s}}$ unlabeled test dataset from domain $\mathcal{D}_{\textrm{s}}$. 
The classical approach to solving the multi-class SVM problem is to consider the problem as a collection of binary classification problems, where the $z$-th classifier constructs a hyperplane between class $z$ and the rest of the $Z-1$ classes. Hence, using a Lagrangian formulation of the classical linearly constrained optimization problem, the final dual problem can be written as \cite{svm}:
\begin{align}
   \underset{\gamma}{\text{max}}&\Bigg\{ \sum_{l=1}^{n_{\textrm{r}}} \gamma_l-\frac{1}{2} \sum_{l=1}^{n_{\textrm{r}}} \sum_{{l'}=1}^{n_{\textrm{r}}}\gamma_l\gamma_{l'}\tau_{\textrm{r}}^l\tau_{\textrm{r}}^{l'}\boldsymbol{K}\bigg(\boldsymbol{\hat{s}}_{\textrm{r}}^l,\boldsymbol{\hat{s}}_{\textrm{r}}^{l'} \bigg)\Bigg\},\label{svm_opt}\\
   & \text{s.t.} \sum_{l=1}^{n_{\textrm{r}}}\tau_{\textrm{r}}^l\gamma_l, 0\leq\gamma_l,\forall l=1,2,\ldots,n_{\textrm{r}},\nonumber
\end{align}
where the kernel matrix $\boldsymbol{K}\bigg(\boldsymbol{\hat{s}}_{\textrm{r}}^l,\boldsymbol{\hat{s}}_{\textrm{r}}^{l'} \bigg)$ is computed by \eqref{inner_product}. 
Then, the decision function for predicting the  transmitted MIMO signal for any test symbol $\boldsymbol{\hat{s}}_{\textrm{s}}^k$ from $\{\mathcal{S}_{\textrm{s}}\}$ is given by \cite{svm},
\begin{algorithm} [h]
\small
\caption{Overview of the proposed GFK G-SVM}
\label{alg_SVM}
\begin{algorithmic}[1]
\State {\textbf{Inputs}\hspace*{2.4mm}\textbf{:} Training dataset \( \{\mathcal{S}_{\textrm{r}},\mathcal{T}_{\textrm{r}}\}\!=\!{\{(\boldsymbol{\hat{s}}_{\textrm{r}}^l,\tau_{\textrm{r}}^l)\}}_{l=1}^{n_{\textrm{r}}},  \)}
\State {\hspace*{12.4mm} \(\boldsymbol{\hat{s}}_{\textrm{r}}^l\in \mathbb{C}^{1\times L'}, \tau_{\textrm{r}}^l \in \{1,2,\ldots,Z\}\);} 
\State {\hspace*{13mm}Test data points \(\{\mathcal{S}_{\textrm{s}}\}={\{\boldsymbol{\hat{s}}_{\textrm{s}}^k\}}_{k=1}^{n_{\textrm{s}}}$, $\boldsymbol{\hat{s}}_{\textrm{s}}^k\in \mathbb{C}^{1\times L'}\);}
\State \textbf{Training:}
\State{Represent the data on \(\mathsf{G}_{L',N}\) through PCA: \(\boldsymbol{S}_{\textrm{r}}, \boldsymbol{S}_{\textrm{s}}\in \mathbb{C}^{L'\times N}\);}
\State{Compute the geodesic flow kernel \(\boldsymbol{F}\) as in \eqref{gfk};}
\State{Compute the kernel matrix \(\boldsymbol{K}\)} by \eqref{inner_product} for \(\mathcal{S}_{\textrm{r}}\) from domain \(\mathcal{D}^{tr}\);
\State{Use the kernel matrix \(\boldsymbol{K}\) in G-SVM algorithm by solving \eqref{svm_opt}};
\State{}
\State \textbf{Classification:}
\State{Return the predicted transmitted MIMO signal using \eqref{solve};}
\end{algorithmic}
\end{algorithm}
\begin{align}
    \mathcal{F} (\boldsymbol{\hat{s}}_{\textrm{s}}^k)=  \Bigg( \sum_{l=1}^{n_{\textrm{SV}}}\gamma_l\tau_{\textrm{r}}^l\boldsymbol{K}\bigg(\boldsymbol{\hat{s}}_{\textrm{r}}^l,\boldsymbol{\hat{s}}_{\textrm{s}}^k\bigg)  + b \Bigg) \label{solve},
\end{align}
where $n_{\textrm{SV}}$ is the total number of support vectors,   $\gamma_l$ are the Lagrange multipliers and $b$ is the bias. 

The MIMO signal detection method based on the proposed GFK G-SVM is summarized in Algorithm~\ref{alg_SVM}. As shown in Algorithm~\ref{alg_SVM}, the geodesic flow kernel is only computed once during the training phase. During the test phase, the algorithm then simply uses the GFK in G-SVM for MIMO signal detection using \eqref{solve}. Hence, no online training is required during testing.  
\section{Simulation Results and Analysis}\label{simulation}
We consider $M=2$ transmit antennas at the mobile node and $N=4$ receive antenna at the receiver. The $4\times 2$ channel is temporally correlated with the speed specified by the normalized Doppler shift $f_{\textrm{D}}T_{\textrm{s}}=0.01$ for a node velocity of $66$ miles per hour (mph), where $T_{\textrm{s}}$ is the baseband sampling interval or time slot. In our simulation, we consider the symbol error rate (SER) as the learning performance. The symbol vectors are generated by the classic mixture of Gaussian (MoG) model \cite{mog} which is widely adopted in the machine learning literature. The number of classes is set as $Z=12$ and the dimension of each data sample is set at $L=48$.  We generate $1,\!000$ random samples for test dataset and all the results presented in this section are averaged over these $1,\!000$ test samples.
\vspace{-0.04in}
\subsection{Simulation Results}
The average SER performance of various MIMO signal detection methods for a target SER performance, i.e, ${10}^{-3}$, with 15 dB SNR is summarized in Table~\ref{tab:MLcompare}. From Table~\ref{tab:MLcompare}, we observe that the DNN based methods in \cite{OAMPnet, HyperMIMO, MMNet} require large training samples in the order of hundreds of thousands to millions. Also, the methods in \cite{OAMPnet, HyperMIMO, MMNet} require online training to achieve their target performance. In contrast, our proposed framework only requires $1,\!200$ training samples and no online training is required. Moreover, the architectures of the DNN-based methods are comprised of large training parameters and many layers, while the proposed method does not have any free learning parameters to train and, hence, does not need online fine-tuning. The above observations suggest that the proposed GFK G-SVM is a promising solution for dynamic settings.
\begin{table}[htbp]
\captionsetup{justification=centering}
\caption{\scriptsize MIMO signal detection schemes for a target SER performance, i.e., ${10}^{-3}$, for 15 dB SNR}
\footnotesize
\centering
\scalebox{0.76}{
\renewcommand{\arraystretch}{2.5}
\begin{tabular}{|l|c|c|c|c|c|c|}
\hline  \parbox{1.9cm}{\textbf{Method}} & \parbox{0.5cm}{\centering \textbf{CSI Used}} & \parbox{1.0cm}{\centering \textbf{Number of\\ Samples}}  & \parbox{1.2cm}{\centering\textbf{Number of\\ parameters}} & \parbox{0.8cm}{\centering\textbf{\centering Number of\\ layers}}  & \parbox{1.05cm}{\centering \textbf{Online Training \\required}} & \textbf{Approach} 
\\
\hline
\parbox[r]{1.9cm}{Proposed\\ GFK G-SVM} & No & $1,200$ & /  & / & No &  GFK based
\\
\hline 
\parbox[r]{1.9cm}{OAMPNet \cite{OAMPnet}} & Yes & $2,50,000$ & $20$ & 10 & Yes & DNN 
\\
\hline
MMNet~\cite{MMNet} & Yes & $500,000$ & $200$ & 10 & Yes & DNN
\\
\hline
\parbox[r]{1.9cm}{HyperMIMO \cite{HyperMIMO}} & Yes & $25,00,000$ & $1,600$ & $8$ &  Yes & DNN
\\ 
\hline 
\end{tabular}
}
\label{tab:MLcompare}
\vspace{-0.15in}
\end{table}
\begin{figure}[htbp]
\centerline{
\includegraphics[width=9cm, height= 6.5cm]{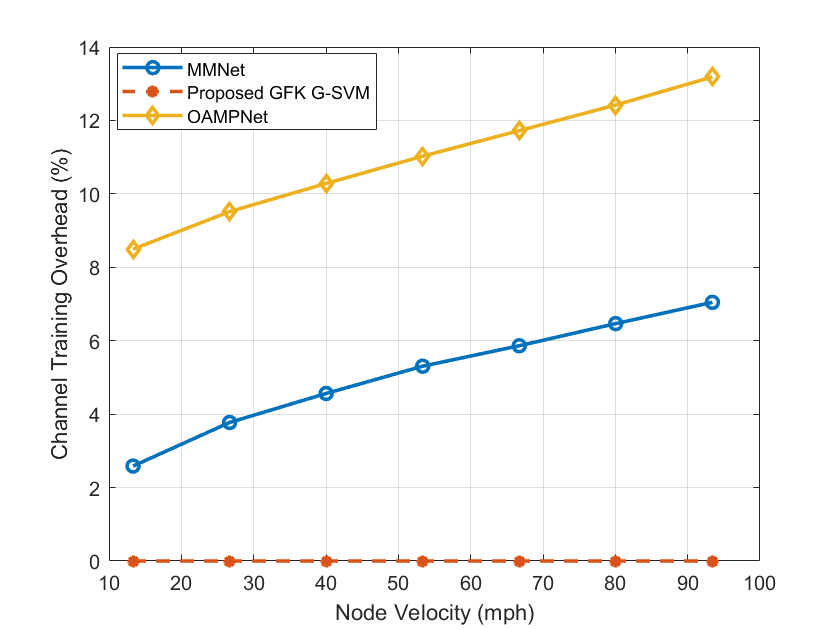}}
\caption{\small{The training overhead for obtaining  CSI and online training versus node velocities for the target SER of ${10}^{-3}$.}}
\label{overhead}
\vspace{-0.08in}
\end{figure}

In Fig.~\ref{overhead}, we compare the online training overhead of various MIMO signal detection methods for different node velocities ranging from $13.33$ mph to $93.42$ mph, corresponding to the Doppler shifts ranging from $0.002$ to $0.014$ for a target SER performance of ${10}^{-3}$. Under such dynamic settings, the DNN based methods in \cite{OAMPnet} and \cite{MMNet} need time to acquire CSI and use it for their online training for adapting to a domain shift. 
Let $P$ be the time required for online training and $F$ be the duration of the data frame. Then, the channel training overhead can be computed as $\frac{P}{P+F}$. From Fig.~\ref{overhead}, we observe that the overhead of OAMPNet~\cite{OAMPnet} and MMNet~\cite{MMNet} increases monotonically with node velocity, as channel fading becomes faster which necessitates more frequent online training for these methods in \cite{OAMPnet} and \cite{MMNet}. In contrast, our proposed framework does not require channel training, thus avoiding this overhead. For instance, for a node velocity of $80.05$ mph (i.e., Doppler shift of $0.012$), the training overhead resulting from OAMPNet~\cite{OAMPnet} and MMNet~\cite{MMNet} is higher by about $12.41\%$ and $6.46\%$, respectively, compared to our proposed framework that does not require channel training.
\begin{figure}[htbp]
\centerline{
\includegraphics[width=9cm, height= 6.5cm]{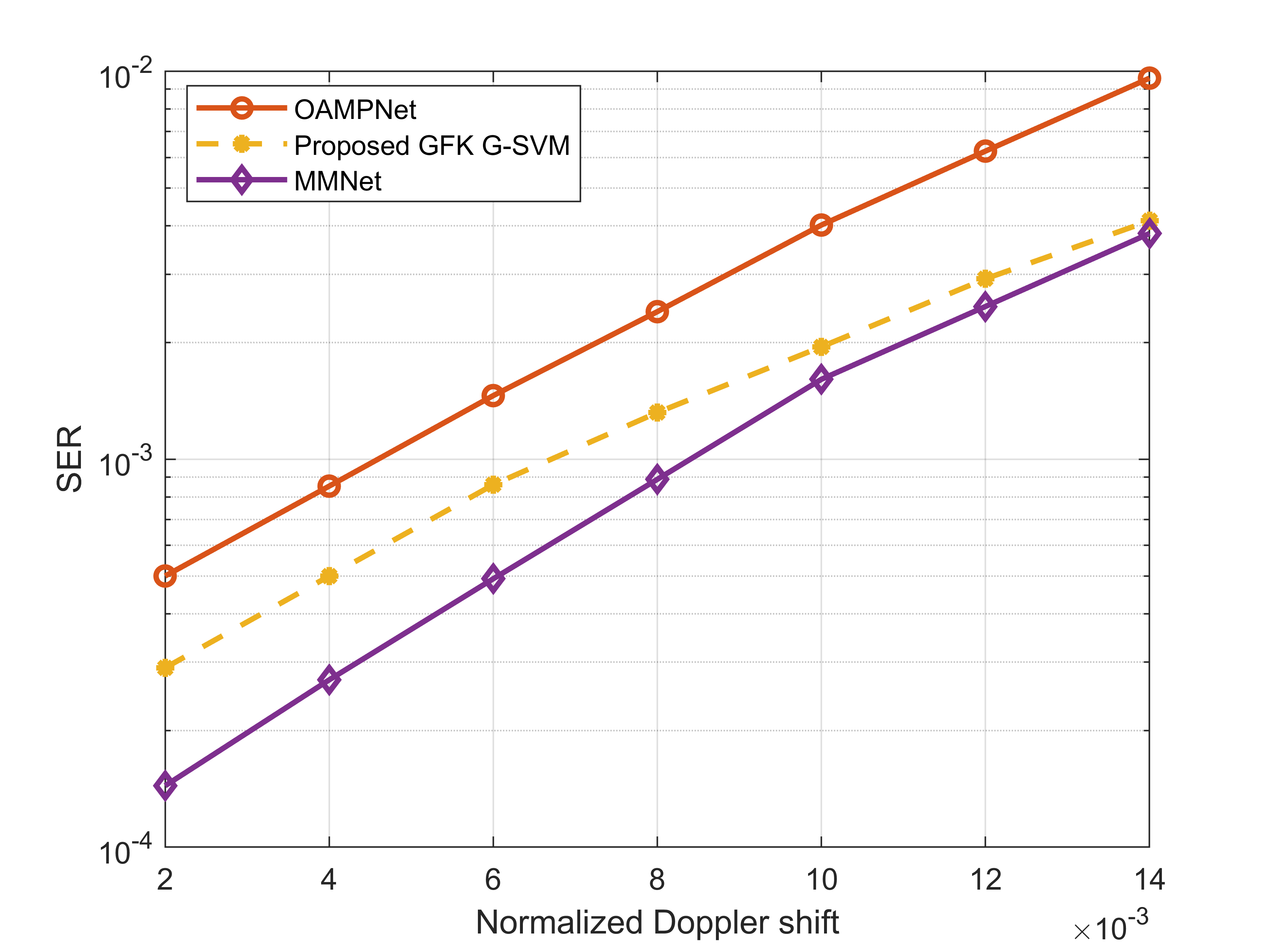}}
\caption{\small{Average SER performance for different Doppler shift with the average transmit SNR equal to 15 dB.}}
\label{doppler}
\vspace{-0.3in}
\end{figure}
\begin{figure}[htbp]
\centerline{
\includegraphics[width=9cm, height= 6.5cm]{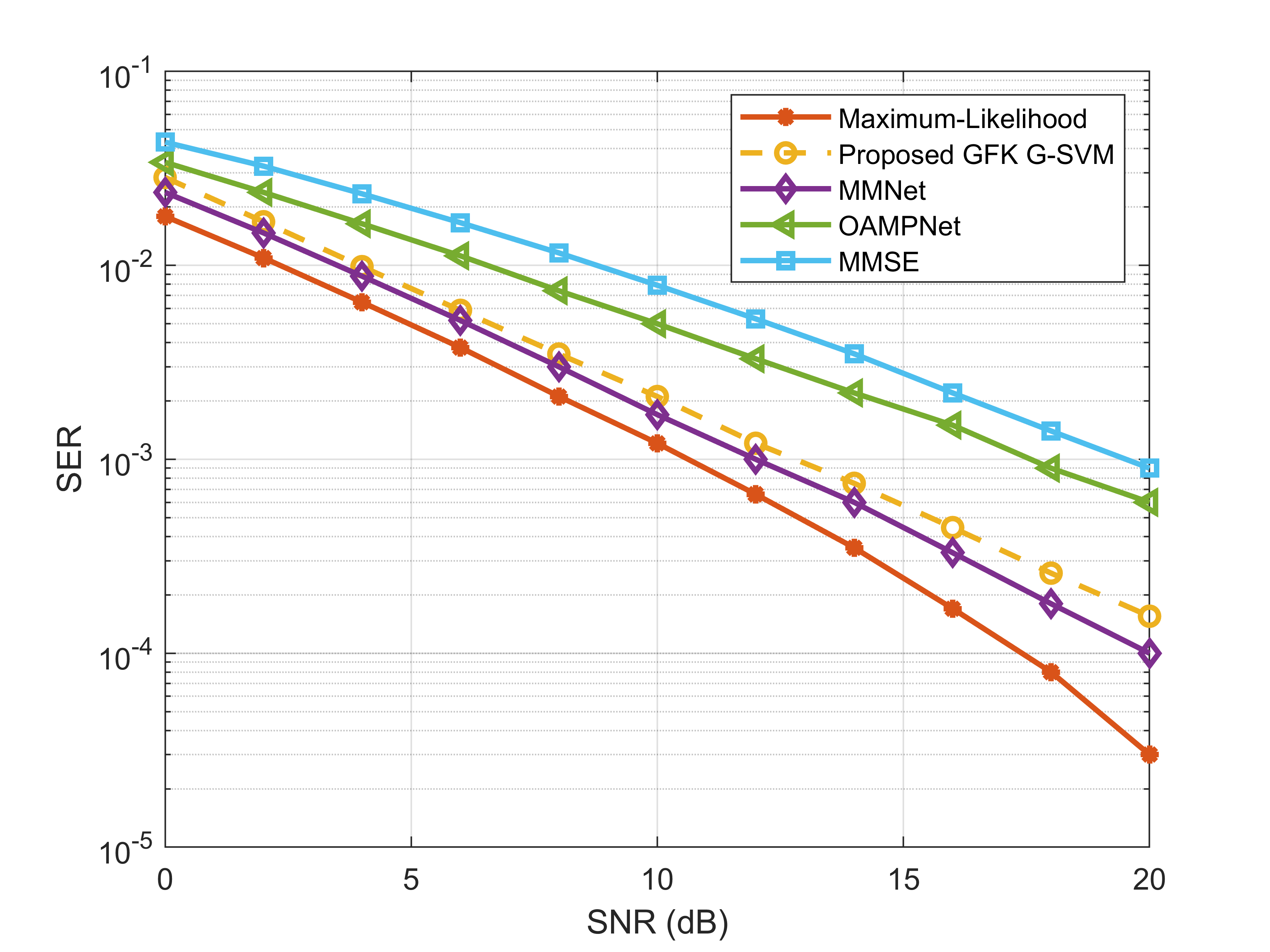}}
\caption{\small{Average SER performance for different average transmit SNR with the normalized Doppler shift fixed at $6\times {10}^{-3}$.}}
\label{SNR}
\vspace{-0.1in}
\end{figure}

Next, in Fig.~\ref{doppler}, we compare how our proposed framework performs under different Doppler shifts without any online training, compared to the DNN based methods in \cite{OAMPnet} and \cite{MMNet} which require online training. In particular, we compare the average SER performance for different normalized Doppler shifts ranging from $0.002$ to $0.014$, corresponding to node velocities ranging from $13.33$ mph to $93.42$ mph with a fixed SNR of $15$~dB. From Fig.~\ref{doppler}, we can see that the performance gap between the proposed method and the  OAMPNet~\cite{OAMPnet} increases in the range of moderate to large Doppler shift (i.e., above 0.008). On the other hand, MMNet~\cite{MMNet} achieves a slightly lower SER, compared to the proposed approach. However, as mentioned, MMNet~\cite{MMNet} requires significant channel training overhead for online training, whereas our proposed framework does not need any online training, and hence, there is no channel training overhead. 
\begin{table*}[htbp]
\caption{\small Computational complexity analysis.}
\centering\scalebox{0.85}{
\renewcommand{\arraystretch}{2.5}
\begin{tabular}{|c|c|c|c|c|c|}
\hline Scheme & MMSE~\cite{ZF} & OAMPNet~\cite{OAMPnet} & MMNet~\cite{MMNet} & Maximum Likelihood \cite{ZF}& Proposed GFK G-SVM
\\
\hline 
\parbox[c]{2.6cm} {\centering Data Features\\Extraction Complexity} & / &  \parbox[c]{2.4cm} {\centering Depends on layers\\and parameters}  & \parbox[c]{3.8cm} {\centering Depends on layers\\and parameters}  & \parbox[c]{3.44cm} {\centering /} & \parbox[c]{3.4cm} {\centering Representation over manifold: $\mathcal{O}({N}^2)$}\\
\hline 
\parbox[c]{2.6cm} {\centering Machine Learning \\ Complexity} & / & Scale as $\mathcal{O}({N}^3)$ & Scale as $\mathcal{O}({N}^2)$ & / &  $\mathcal{O}({N})+\mathcal{O}(n_{SV}{N})$\\ 
\hline 
Total Complexity & $\mathcal{O}({N}^3)$  &  $\mathcal{O}({N}^3)$  & \parbox[c]{3.8cm} {\centering $\mathcal{O}({N}^2)$ } &  \parbox[c]{3.4cm} {\centering $\mathcal{O}({N}^3)$ }& \parbox[c]{3.4cm} {\centering $\mathcal{O}({N}^2)$}\\
\hline
\end{tabular}
}
\label{complexity_compare}
\vspace{-0.05in}
\end{table*}

Finally, in Fig.~\ref{SNR}, we compare the average SER performance as a function of the SNR and with fixed normalized Doppler shift $0.006$. From Fig.~\ref{SNR}, we can see that the proposed method outperforms both the DNN based OAMPNet \cite{OAMPnet} and the traditional MMSE~\cite{ZF} approach with a large margin, reaching up to $82.75\%$ and $74.13\%$  better SER performance at $20$ dB SNR than the MMSE~\cite{ZF} and OAMPNet \cite{OAMPnet}, respectively. The performance is also comparable to MMNet \cite{MMNet} and the optimal maximum-likelihood detectors \cite {ZF}. However, as mentioned, MMNet \cite{MMNet} requires a large volume of training samples as well as online training, whereas our proposed framework requires much fewer training samples and does not need online training. Moreover, the optimal maximum-likelihood detectors are highly computationally complex and require perfect CSI, which is extremely hard to get in such dynamic settings.

\subsection{Computational Complexity Analysis}
We next analyze the computational complexity of our proposed framework and compare it with other benchmark algorithms, as summarized in Table \ref{complexity_compare}. Our proposed approach is implemented in two steps: 1) computing the geodesic flow kernel over the Grassmannian manifolds, and 2) performing classification for MIMO signal detection.  The computational complexity for computing the geodesic flow kernel over Grassmannain manifolds is  $\mathcal{O}({N}^2)$. Meanwhile, the computational complexity for G-SVM classification can be computed as 
\begin{align}
   \mathcal{O}(n^2Z{N}) +\mathcal{O}(n_{\textrm{SV}}{N}) \approx \mathcal{O}({N}), 
\end{align}
where $n$ is the number of training samples, $Z$ is the number of classes, $n_{\textrm{SV}}$ is the total number of support vectors, and $N$ is the dimension of the subspace. Here, $n$, $Z$, and $N$ are fixed for the proposed scheme. Thus, the total computational complexity of the proposed GFK G-SVM method scales as $\mathcal{O}({N}^2)$, which is similar to the MMNet \cite{MMNet}.  However, unlike MMNet \cite{MMNet}, it does not require a large training sample and online training. Moreover, the computational complexity of our proposed framework is lower than MMSE \cite{ZF}, maximum-likelihood \cite{ZF} and OAMPNet \cite{OAMPnet},
as summarized in Table \ref{complexity_compare}.    

\balance
\section{Conclusion} \label{conclusion}
In this paper, we have addressed the challenge of MIMO signal detection under rapid domain shifts in dynamic wireless communication settings. Towards this goal, we have proposed a novel GFK based MIMO signal detection method over Grassmannian manifolds. We have first represented the received MIMO signals 
as points on Grassmannian manifolds by formulating basis of subspaces spanned by the rows vectors of the received signal matrices. Then we have computed the GFK over the subspaces of the training and test data domains. As such, we have integrated an infinite number of subspaces that lie on the geodesic flow from the training subspace to the test one through GFK. We have shown that the flow along the geodesic represents incremental changes in geometric and statistical properties of the received signals due to mobility. By learning from all of these changes, the proposed GFK kernel can extract those representations that are invariant to the domain shift. Subsequently, we have used the kernel in G-SVM for MIMO signal detection. Simulation results show that the proposed method achieves competitive SER performance against the existing baselines like  OAMPNet \cite{OAMPnet}, HyperMIMO \cite{HyperMIMO} and MMNet\cite{MMNet} with
only $1,\!200$ training samples and without online retraining.
\balance

\bibliographystyle{IEEEbib}

\bibliography{Domain_Adapt_refs}

\end{document}